\documentclass[german,english]{bvm2020}

\usepackage{tikz}
\usepackage{graphicx}
\usepackage[misc]{ifsym}
\usepackage{multirow}
\usepackage{adjustbox}
\definecolor{siemens}{RGB}{10, 157, 157}
\definecolor{health}{RGB}{243, 117, 0}
\def\articlenumber{0000}

\date{}
\title{Deep learning-based denoising of mammographic images using physics-driven data augmentation}

\titlerunning{Deep learning-based denoising of mammographic images}
\author{Dominik Eckert\inst{1}(\Letter), Sulaiman Vesal\inst{1}, Ludwig Ritschl\inst{2}, Steffen Kappler\inst{2}, Andreas Maier\inst{1}}

\authorrunning{Eckert et al.}
\institute{Pattern Recognition Lab, Friedrich-Alexander-Universit\"at Erlangen-N\"urnberg, Germany\\
\and Siemens Healthcare GmbH, Forchheim, Germany\\}

\email{dominik.17.eckert@fau.de}

\begin{document}
%==============================================================================
% wählen Sie mit dem Befehl \selectlanguage die Sprache aus, in der Ihr 
% Proceeding verfasst ist
%
%\selectlanguage{german}
\selectlanguage{english}

\maketitle

\begin{abstract}

Mammography is using low-energy X-rays to screen the human breast and is utilized by radiologists to detect breast cancer. Typically radiologists require a mammogram with impeccable image quality for an accurate diagnosis.
In this study, we propose a deep learning method based on Convolutional Neural Networks (CNNs) for mammogram denoising to improve the image quality. We first enhance the noise level and employ Anscombe Transformation (AT) to transform Poisson noise to white Gaussian noise. With this data augmentation, a deep residual network is trained to learn the noise map of the noisy images. 
We show, that the proposed method can remove not only simulated but also real noise. Furthermore, we also compare our results with state-of-the-art denoising methods, such as BM3D and DNCNN. In an early investigation, we achieved qualitatively better mammogram denoising results.
\end{abstract}

\section{Introduction}
According to the World Health Organization (WHO), cancer is the second leading cause of death globally and is responsible for 9.6 million deaths in 2018 \cite{globocan2018}. Among these, breast cancer is the leading cause of women's deaths and accounts for 15 percent of its \cite{globocan2018}.

One of the most widely used modalities for breast cancer detection is mammography. It is using low-energy X-rays to screen the human breast and helps the radiologist to detect breast cancer in an early stage. For an accurate diagnosis, impeccable image quality is required. This is due to the complexity of mammograms, and the small size of microcalcifications, which are essential for breast cancer detection. Denoising the mammogram is one approach to improve quality.  Unlike in the research field of computer vision, few attempts have been made to use convolutional neural networks (CNN's) to reduce the noise in mammograms \cite{cnn_challenges_mamo_2019}\cite{mammo_denoising_survey_2017}. In \cite{deepGurpem2018}, they used a CNN with 17 layers and the MIAS-mini(MMM) data set for training. To date, we are not aware of works aiming at the correct physical modelling of noise reduction in deep learning following ideas the ones presented in \cite{Maier19-MI}. %On the contrary, we train the network on images without preprocessing. That allows modeling of physically motivated noise. 

In this paper, we investigate the feasibility of CNNs for denoising mammograms. This does not only include the development and training of an adequate network, but also the examination of detail preservation since the latter one is crucial for breast cancer detection. We propose a deep residual network, which could denoise the mammogram images with high accuracy.

\section{Material and Methods}

Because of the quantum nature of light, photons arrive at random times. %\cite{Lo1979}
This leads to an uncertainty about the received signal, which can be modeled with a Poisson distribution,
where each pixel value $ z $ depends on the arrival rate $ \lambda $ of the photons. The distribution can be described as the following: %\cite{poisson1837recherches}
\begin{equation}
P(z|\lambda) = \frac{\lambda^{z} e^{-\lambda}}{z!}
\end{equation}
Therefore the detector receives a noisy X-ray image $ Y $, which can be decomposed in a noise-free signal $ \mu(Z) $ and the noise $ V $, so that $Y = \mu(Z) + V$ holds. Where $\mu(Z_{i,j})$ is the mean of a pixel value.
This study aims to train a network to estimate the noise $ V $, by feeding it the noisy image $ Y $ \cite{beyondZhang2016}.

\subsection{Data and Augmentation}

\begin{figure}[tb]
    \centering
        \includegraphics[width=0.7\textwidth]{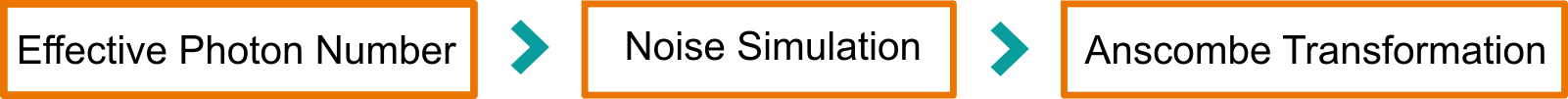}
    \caption{Data augmentation pipeline for FFDM images.}
    \label{fig:test}
\end{figure}

We use a data set of 125 Full-Field Digital Mammograms (FFDMs), which was provided by our industry partner.
The X-ray images are prepared as shown in Fig. \ref{fig:test}. In the first step, the real number of photons is calculated out of the pixel values.
In the second step, a dose reduction is simulated by generation of more noise, followed by a transformation of the signal-dependent Poisson noise using the Anscombe transformation to white Gaussian noise \cite{Anscombe1948}. To model the Poisson noise reasonably,  we exploit the linear dependency of pixel value $ z $ of the real number of photons $ \lambda $ by a factor $ k $  to get $ \lambda $.
It employs, that $ z = k \cdot \lambda $ holds. Since $k$ is unknown, it has to be obtained, by dividing the variance of $ z $ over the mean of $ z $:
\begin{equation}
    \frac{\text{var}(z)}{\text{mean}(z)} = \frac{k^2 \cdot \text{var}(\lambda)}{k \cdot \text{mean}(\lambda)} = k
\end{equation}
To simulate more noise on the ground truth images, the arrival rate of $ \lambda $ of the photons will be scaled down by a factor $ \alpha $ using the following equation:
\begin{equation}
	P(Z| \alpha \lambda ) = \frac{\alpha \lambda^{Z} \cdot e^{- \alpha \lambda}}{Z!}
\end{equation}
This results in a new mean  $ \mu = \alpha \lambda $  and variance $ \sigma^2 = \alpha \lambda $.
Now the Anscombe transformation can be applied to transform the signal-dependent Poisson noise to signal independent white Gaussian noise with a standard deviation of one \cite{Anscombe1948}:
\begin{equation}
A: z \to 2 \sqrt{z + \frac{3}{8}}
\end{equation}

\subsection{Network Architecture}
\begin{figure}[tb]
	\setlength{\figbreite}{0.3\textwidth}
	\centering
	\subfigure[Residual Network]{\includegraphics[width=\figbreite]{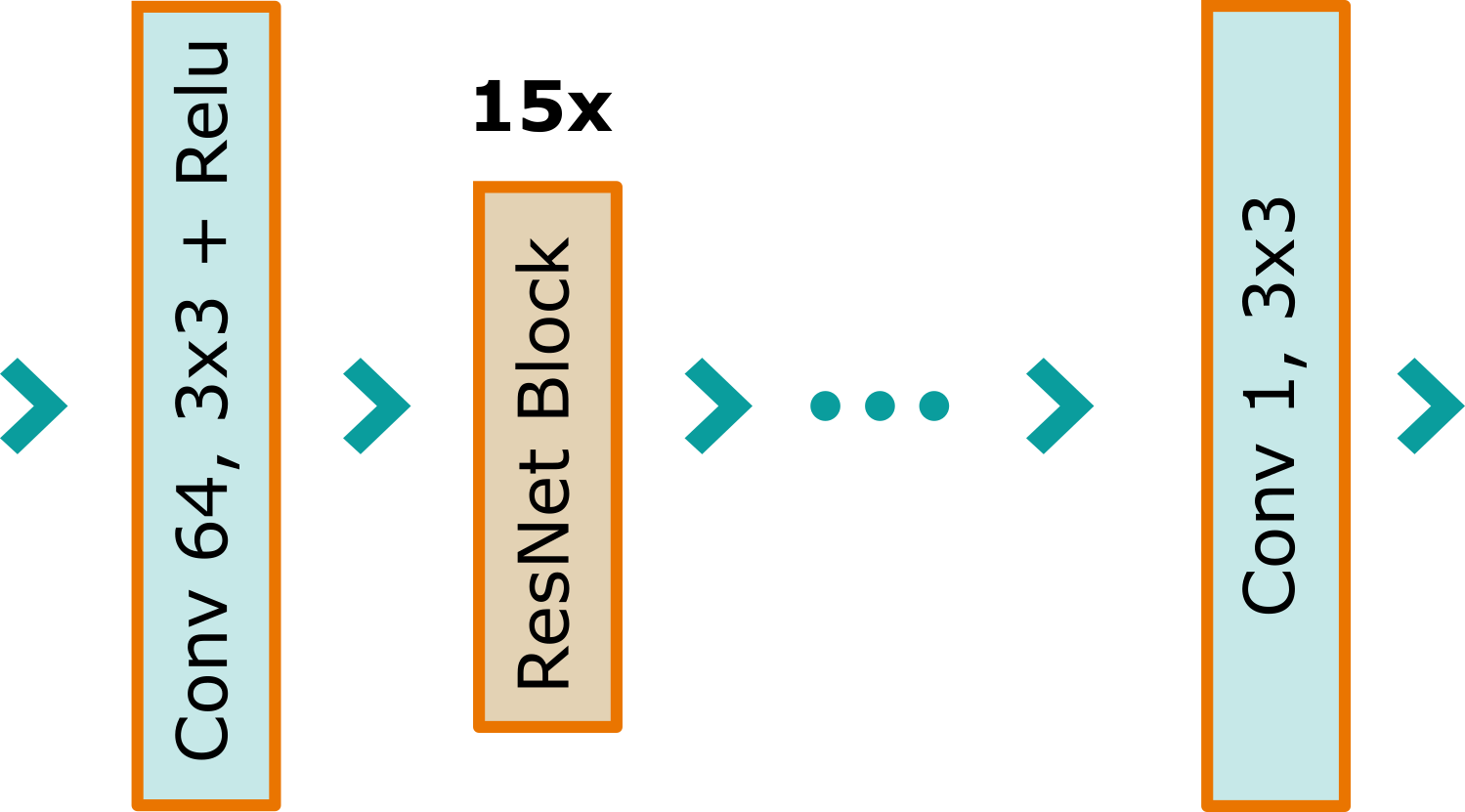}}
	\subfigure[Residual Block]{\includegraphics[width=\figbreite]{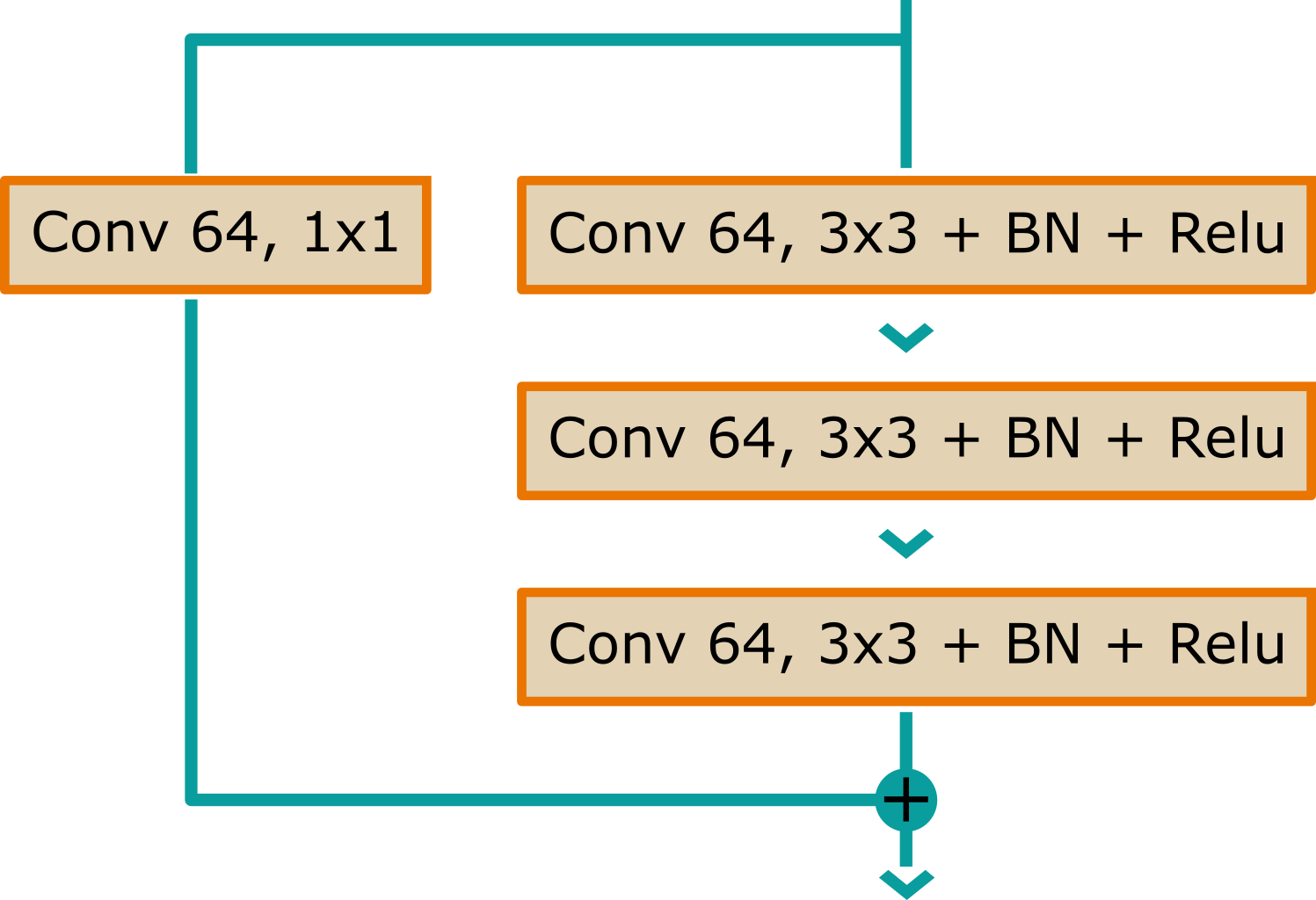}}
	\caption{Residual Network Architecture.}
	\label{fig:resnet}
\end{figure}

We use a residual network, as shown in Fig. \ref{fig:resnet}. The first layer consists of a convolution layer with 64 kernels of size $3\times3$ and a ReLU activation function.
The middle part of the network consists of fifteen residual blocks. Each block is composed of three convolution layers with 64 kernels of the size of $3\times3$ and followed by a Batch Normalization and a ReLu. We choose 3 convolution layers in a row for each residual block, to ensure that the information contained in the receptive field is high enough for each residual block. The same padding is used for all layers, and the skip connection can be molded by convolution with 64 Kernels of size $1\times1$. This setup leads to a network with 1.700.000 trainable parameters.

%\subsection{Loss Function}
\textbf{Loss Function.} For denoising with deep learning, often a Mean Square Error (MSE) is used as a loss function. It is defined as $ \text{MSE} = \frac{1}{n}\sum_i (V_i - \hat{V_i})^2 $,
where $ V $ is the true and $ \hat{V} $ the estimated noise.
As argued by Zhao et al. \cite{zhao2016loss}, MSE is not the best choice to use as a loss function, because of its difference to human perception.
For this reason, they use a combination of the Structural Similarity Index (SSIM) with MSE.
The SSIM $S(x,y) = f \left( l(\mathbf{x},\mathbf{y}), c(\mathbf{x},\mathbf{y}), s(\mathbf{x},\mathbf{y}) \right)$  is formed by three different equations:
$ l(\mathbf{x}, \mathbf{y}) $ measures similarity of the luminance, $  c(\mathbf{x},\mathbf{y}) $ the difference of the contrast
and $  s(\mathbf{x},\mathbf{y}) $ compares the structures \cite{wang2004image}. We also use a combination of MSE combined with SSIM. Because SSIM is 1 if the compared images are the same, and gets smaller if they differ, we have to use $ 1 - \text{SSIM}$ in the loss. This term is usually ten times smaller than the MSE. To weight both metrics equally, we formulate the loss like $ \text{L}_\text{total} =  \text{MSE} + 10 \cdot (1 - \text{SSIM}) $.

\textbf{Training.} For training, the X-ray images are cropped to patches of size 64$\times$64. We used 100 images for training and 25 for validation which leads to 40.000 patches for training and 10.000 patches for validation. For training the network, we use Adam optimizer with a learning rate of 0.0001. The noise is simulated by the means of a dose reduction of $ 80 \% $. To ensure that the network never sees the same noise twice, the noise is simulated at each iteration. The training of the network takes 125 epochs before it is stopped.

\section{Experiments and Results}
\subsection{Denoising of simulated noise and performance comparison}
\begin{figure}[tb]
\subcapcentertrue
\centering
\setlength{\figbreite}{0.19\textwidth}
\subfigure[FFDM]{\includegraphics[width=\figbreite]{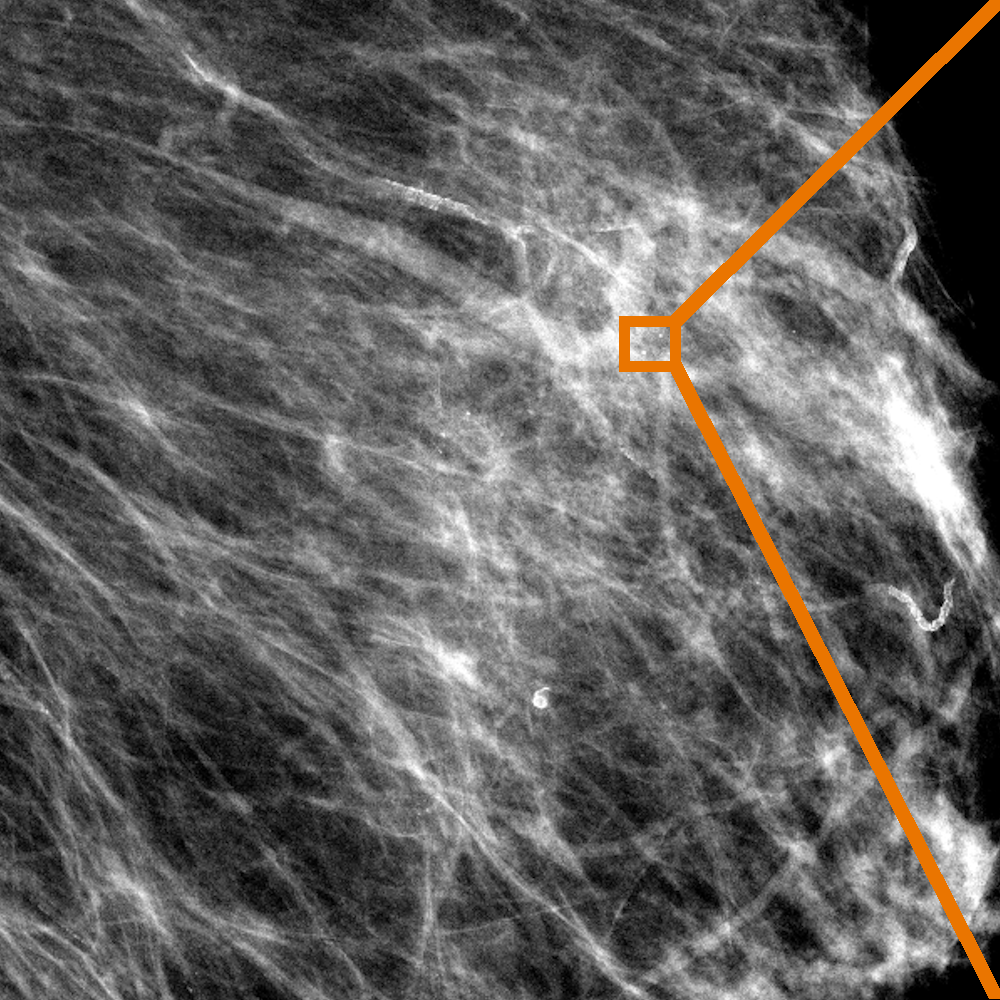}}
\subfigure[Noisy]{\includegraphics[width=\figbreite]{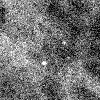}}
\subfigure[GT]{\includegraphics[width=\figbreite]{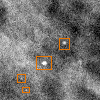}}
\subfigure[Gaussian]{\includegraphics[width=\figbreite]{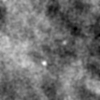}}

\subfigure[BM3D]{\includegraphics[width=\figbreite]{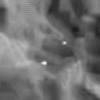}}
\subfigure[Gurprem]{\includegraphics[width=\figbreite]{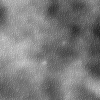}}
\subfigure[DNCNN]{\includegraphics[width=\figbreite]{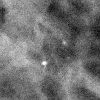}}
\subfigure[ResNet]{\includegraphics[width=\figbreite]{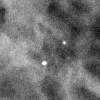}}
\subfigure[Diff: (h)-(c)]{\includegraphics[width=\figbreite]{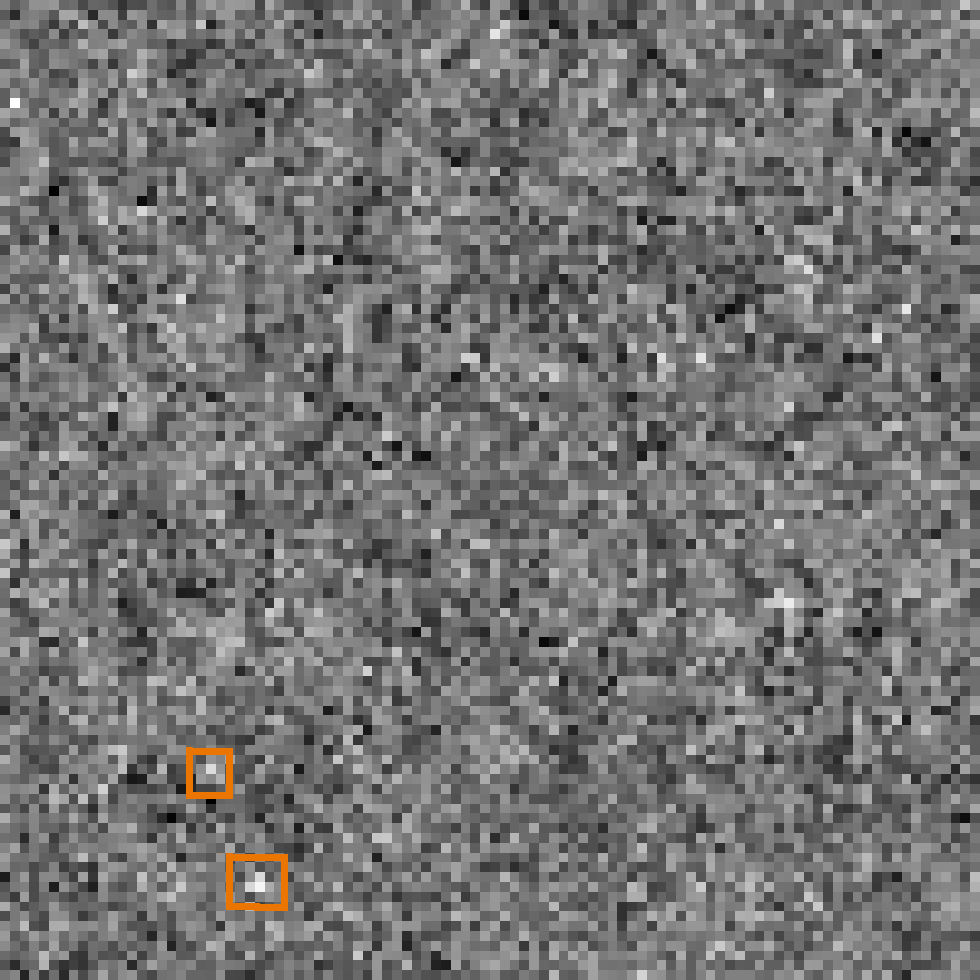}}
\caption{Results of different denoising methods. Microcalcifications are marked with an orange square in the GT and the difference image. Two are still clearly visible in the results of the ResNet. The smaller ones are not distinguished from noise by the ResNet and therefore appear in the difference image. They are also not detectable for the human eye in the noisy image.}
\label{fig:comp}
\end{figure}

We evaluated the denoising capacity of our residual network on X-ray images with a simulated dose reduction of $ 80 \% $. The denoised images are shown in Fig. \ref{fig:comp}. In Table \ref{tab:comp}, we show the PSNR, SSIM and the standard deviation ($ \sigma_{\text{image}} $) of the different outputs of the denoising methods. We also indicate the $ \sigma_{\text{image}} $, because the noise reduction goes below the noise level of the GT. Therefore, one should consider that MSE and PSNR do not reflect entirely the noise reduction anymore. For a better visibility, we present the results of a small area of the original image, which contains two microcalcifications. 

The result of the residual network compared against a Gaussian filter with $ \sigma = 1$ and the state-of-the-art algorithm BM3D \cite{dabov2007image}. For both algorithms, we perform the same noise simulation and make use of the Anscombe transformation. 

In Gurprem et al. \cite{deepGurpem2018}, they trained a CNN with 17 layers similar to Zhang et al. \cite{beyondZhang2016}. They downsample the training images to a size of  $512 \times 512$ and simulate Gaussian noise with $\sigma = 25$.  Initially, we retrained this network similar to Gurprem et al., but instead of scaling down our images of size $ 3518 \times 2800$, we use patches of size $512 \times 512$ for training. In a second attempt, we retrained the 17 layer CNN of Zhang et al. \cite{beyondZhang2016} with our method. This includes Poisson noise, Anscombe transformation, and MSE + SSIM loss and refers to it as DNCNN.

% We also retrained a network, like it is down in Gurprem et al. \cite{deepGurpem2018}. But we train it with our data set and instead of scaling down our images of size $ 3518 \times 2800$ to a size of $512 \times 512$, we trained with patches of size $512 \times 512$. Like in Gurprem et al. Gaussian noise with $\sigma = 25$ is used for training. They also use a Convolutional network with 17 layers, like it was proposed in Zhang et al. \cite{beyondZhang2016}. In a second attempt, we use this architecture and retrain this network with our methods, that is Poisson noise, Anscombe transformation, and MSE + SSIM loss and refer to it as DNCNN. 

The $ \sigma_{\text{image}} $ of the GT Image is 0.042. All methods produce results with smaller $ \sigma_{\text{image}} $ than the GT.  However, the results of the Gaussian filter and DNCNN have the highest $ \sigma_{\text{image}} $ value (0.41), and the Gaussian filter tends to blur the image (Fig. \ref{fig:comp}d). This is not the case with the BM3D model (Fig. \ref{fig:comp}e). Alternatively, the output is over smoothed and artifacts introduced, but microcalcifications are still visible. Fig. \ref{fig:comp} (f) shows the output,  when employing the trained network of Gurprem et al. It can be seen that microcalcifications are almost lost because of image averaging. We assume, that a noise level of $\sigma  = 25$ is extremely high for training images of our data set. Because of their wide intensity range, many details are suppressed with that noise level. Consequently, the network is not able to learn detailed features anymore. The same network trained with our method produces satisfactory results but reduces the visibility of some microcalcifications (Refers to Fig. \ref{fig:comp}h). Using our proposed network and methods, microcalcifications are still clearly visible and no additional artifacts are introduced. This is also reflected in the highest PSNR and SSIM of  $ 36.18 $ and $ 0.841 $, respectively. In the last image (Fig. \ref{fig:comp}a), we show the difference between the GT and the denoised image with ResNet. One can see those small microcalcifications appear in the left lower corner of the difference image. Hence, the network has difficulties in distinguishing them from noise.

% Table generated by Excel2LaTeX from sheet 'Sheet1'
 \begin{table}[tb]
   \centering
   \caption{Performance comparison between our proposed method and other denoising methods.}
     \begin{tabular}{lccc}
     \hline
     \textbf{Methods} & \textbf{PSNR} & \textbf{SSIM} & \textbf{$\sigma_{\text{image}}$} \\
     \hline
     \hline
     Noisy                                          & 30.94  & 0.64 & 0.057 \\
     Gaussian filter                                & 35.48 & 0.815 & 0.041 \\ %\cite{ito2000gaussian} 
     BM3D \cite{dabov2007image}                     & 35.49 & 0.801 & \textbf{0.038} \\
     Gurprem et al. \cite{deepGurpem2018}           & 33.50 & 0.781 & \textbf{0.038} \\
     DNCNN  \cite{beyondZhang2016}                      & 36.08 & 0.840 & 0.041 \\
     ResNet (ours)                                  & \textbf{36.18} & \textbf{0.841} & 0.040 \\
     \hline
     \end{tabular}%
   \label{tab:comp}%
 \end{table}%

\subsection{Denoising real-world noise}
\begin{figure}[tb]
\centering
\setlength{\figbreite}{0.4\textwidth}
\subfigure[Low Dose FFDM, $ \sigma_{\text{image}} = 0.0442 $]{\includegraphics[width=\figbreite]{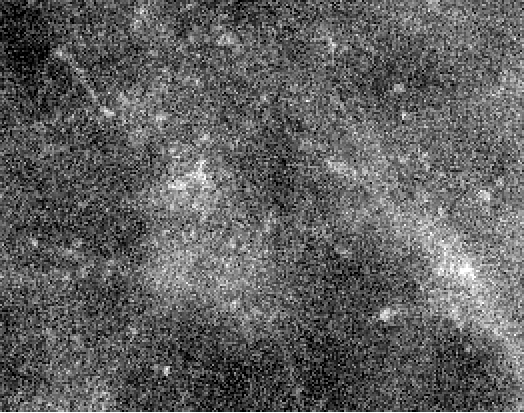}}
\subfigure[Denoised FFDM, $ \sigma_{\text{image}} = 0.0302$]{\includegraphics[width=\figbreite]{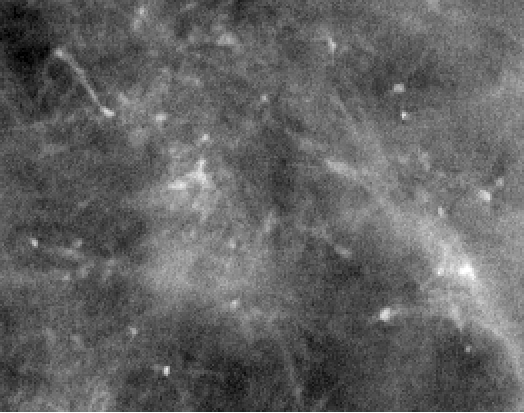}}
\caption{Denoising on real-world noise.}
\label{fig:results}
\end{figure}

We further evaluate the denoising capability of our proposed method on real-world noise. Fig. \ref{fig:results} demonstrate the noisy and the denoised FFDM image. It appears that the network denoise the real noise as good as simulated noise. Similar to simulated noise, microcalcifications are visible, and no additional artifacts introduced in the images. The $ \sigma_{\text{image}} $ value is also significantly low in compare to actual low dose FFDM.

\section{Conclusion}
In this paper, we introduced a deep learning-based method for mammogram denoising, which significantly enhanced the image quality. Our proposed method outperformed state-of-the-art methods such as BM3D, DNCNN, and Gurprem et al. This was shown on a visual example and proved by comparing the PSNR and SSIM values with other methods. However, information which is lost in the noise, can not be restored by the network. One should be careful, by reducing the dose, especially if a high degree of detail is demanded. 

%\noindent\footnotesize{\textbf{Disclaimer:}
%The presented method is not commercially available. Due to regulatory reasons, its future availability cannot be guaranteed.}
\noindent\textbf{Disclaimer:} The concepts and information presented in this paper are based on research and are not commercially available.

\bibliographystyle{splncs04}
\bibliography{biblogr}
% Bitte setzen Sie hier Ihre Beitragsnummer ein und benennen Sie
% die BibTeX-Datei ebenfalls auf Ihre Beitragsnummer um.
%Kontrollzeiledef
\marginpar{\color{white}E\articlenumber} % Zeile nicht verändern!
\end{document}